\documentstyle[epsf,amsfonts,amssymb,eqsecnum,aps,preprint]{revtex}
\tightenlines
\begin{document}
\draft
\title{Polymer depletion interaction between two parallel repulsive walls}
\author{F. Schlesener,$^{1,2}$ A. Hanke,$^3$ R. Klimpel,$^1$ and S. Dietrich$^{2,4}$}
\address{$^1$Fachbereich Physik, Bergische Universit\"at Wuppertal,
D-42097 Wuppertal, Germany}
\address{$^2$Max-Planck-Institut f\"ur Metallforschung,
Heisenbergstr. 1, D-70569 Stuttgart, Germany}
\address{$^3$Department of Physics, Massachusetts Institute of
Technology, Cambridge, MA 02139, USA}
\address{$^4$Institut f\"ur Theoretische und Angewandte Physik,
Universit\"at Stuttgart, D-70550 Stuttgart, Germany}
\maketitle
\sloppy
\begin{abstract}
The depletion interaction between two parallel repulsive walls 
confining a dilute solution of long and flexible polymer 
chains is studied by field-theoretic methods.
Special attention is paid to self-avoidance between 
chain monomers relevant for polymers in a good solvent.
Our direct approach avoids the mapping of the actual polymer chains on
effective hard or soft spheres. 
We compare our results with recent Monte Carlo simulations
[A. Milchev and K. Binder, Eur. Phys. J. B {\bf 3}, 477 (1998)]
and with experimental results for the depletion interaction
between a spherical colloidal particle and a planar wall in a dilute
solution of nonionic
polymers [D. Rudhardt, C. Bechinger, and P. Leiderer, 
Phys. Rev. Lett. {\bf 81}, 1330 (1998)].
\end{abstract}
\bigskip
\narrowtext
\section{Introduction}
\label{intro}
In polymer solutions the overlap of depletion zones for monomers
due to repulsive confining walls or mesoscopic particles
dissolved in the solution induces an important and tunable 
effective interaction potential \cite{konstanz}.
For example, this depletion interaction explains successfully 
phase diagrams of colloid-polymer mixtures 
\cite{sperry,lekkerkerker,poon}.
Recent experimental techniques
facilitate even the measurement of the depletion force between 
a wall and a {\em single\/} colloidal particle 
\cite{ohshima,RBL98,verma}. In the context of such solutions confined
to thin films and porous materials
the geometry of two parallel walls
has been extensively studied as a paradigmatic case
\cite{reviews,unirel,cifra,hagita,kumar,hooper,milchev}.

For strongly overlapping chains as realized in a 
{\em semidilute} polymer solution, chain flexibility is taken 
into account within self-consistent field theory or 
within the framework of phenomenological scaling theory 
\cite{joanny1,gennes1,odijk}. 
On the other hand, in a {\em dilute} polymer solution different chains 
do not overlap so that the behavior of the polymer solution 
is determined by the behavior of a single chain. To a certain extent
and under certain circumstances, 
a single chain can be modeled by a random walk without self-avoidance 
(ideal chain). In three dimensions this situation is closely realized in a 
 so-called $\theta$-solvent \cite{gennes2}. 
If the solvent temperature is below the $\theta$-point 
(poor solvent) the polymer coils tend to collapse
\cite{cordeiro,singh}. However, in the common case that  
the solvent temperature is above the $\theta$-point (good solvent)
the excluded volume (EV) interaction between chain monomers becomes 
relevant so that the polymer coils are less compact than the
corresponding ideal 
chains. The emphasis in this work is on the latter situation and we
investigate the effect of the EV interaction on the depletion
interaction between two parallel walls as compared to the case of
confined ideal chains.

By focusing on {\em long} flexible chains in a system of
mesoscopic size we obtain mostly {\em universal} results which are
independent of microscopic details 
\cite{gennes2,cloizeaux,schafer,eisen,ehd,HED,bringer}.
Due to the universality of the corresponding properties
it is sufficient to choose a simple model for calculating 
these results.
For our investigations we use an Edwards-type model 
\cite{gennes2,cloizeaux,schafer} for the polymer chain 
which allows for an expansion in terms of the EV interaction
and which is amenable to a field-theoretical
treatment via the polymer magnet analogy. 
The basic elements in this expansion are 
partition functions ${\cal Z}^{[0]}_{\rm seg}({\bf r}, {\bf r}')$ for chain
segments which have no EV interaction (as indicated by the superscript $[0]$)
and with the two ends of the segment fixed at ${\bf r}$ and ${\bf r}'$. 
This perturbative treatment has to be carried out in presence of confining
geometries.
We consider two structureless parallel walls in $d$ dimensions and 
a distance $D$ apart
so that in coordinates ${\bf r} = ({\bf r}_{\parallel}, z)$ 
the surface of the bottom wall is located at $z = 0$ 
and ${\bf r}_{\parallel}$ comprises the $d-1$ components of ${\bf r}$
parallel to the walls. The surface of the upper wall is located at
$z = D$. The interaction of the polymer with the non-adsorbing walls
is implemented by the boundary condition that
the segment partition function and thus the partition function for
the whole chain vanishes
as any segment approaches the surface of
the walls \cite{gennes2,eisen}, i.e.,
\begin{eqnarray}
\label{zzero}
{\cal Z}^{[0]}_{\rm seg}({\bf r},{\bf r}') & \to & 0 \, \, ,
\quad z, z'\to 0, D \, \, .
\end{eqnarray}
For the present purpose the only relevant property which characterizes
one of the polymer 
chains is its mean square end-to-end distance in the bulk solution
which we denote by $d \, {\cal R}_x^2$ \cite{ehd,HED,bringer};
for convenience we include the spatial dimension $d$ as a prefactor.
The results presented in the following are obtained for $d=3$ both for
ideal chains and for chains with EV interaction. 
In Sec.\,\ref{sec_ft} we present our results for the interaction 
potential and the force between two parallel walls. We note 
that for chains with EV interaction these results are only valid 
in the limit $D \gg {\cal R}_x$ because our theoretical approach 
is not capable to describe the dimensional 
crossover to the behavior of a quasi 
$d-1$ dimensional system which arises for $D \ll {\cal R}_x$.
In Sec.\,\ref{sec_sim} we compare these results with the simulation 
data of Milchev and Binder 
\cite{milchev}. In Sec.\,\ref{sec_exp} we apply the Derjaguin 
approximation in order to obtain 
from the results in Sec.\,\ref{sec_ft} 
the depletion interaction between
a spherical particle and a wall in a dilute polymer solution 
and compare it with the corresponding experimental results of 
Rudhardt, Bechinger, and Leiderer \cite{RBL98}.

In view of the complexity of the actual experimental systems involving
spatially confined colloidal suspensions dissolved in a solution
containing polymers, in the past the corresponding theoretical
descriptions relied on suitable coarse-grained, effective models
and on integrating out less relevant degrees of freedom.
The gross features of these systems can be obtained by mapping the
polymers onto effective hard spheres as pioneered by Asakura and
Oosawa \cite{asakura}. In a more refined description Louis et al.
\cite{eurich,hans00} derived effective interaction potentials
between polymer coils such that they behave like soft spheres.
This approach allows one to capture the crossover in structural
properties to semidilute and dense polymer solutions. Based on
such a model Louis et al. \cite{hans00} have calculated, inter alia,
the corresponding depletion energy between two parallel repulsive
plates. Besides presenting the depletion energy for ideal chains
in terms of an expansion introduced by Asakura and Oosawa,
they find the occurrence of repulsive depletion forces in an
intermediate regime of $D$ upon significantly increasing the
polymer density. In our present completely analytic study we
focus on dilute polymer solutions, for which the depletion
forces turn out to be always attractive, by fully taking into
account the flexibility of the polymer chains and the
self-avoidance of the polymer segments with a particular
emphasis on long chains. This complementary point of view
allows us  to make contact
to the Monte Carlo simulation data in Ref.\,\cite{milchev} and with
the experimental data in Ref.\,\cite{RBL98}; both these comparisons
have not been carried out before.

\section{Effective interaction between parallel walls}
\label{sec_ft}
\subsection{Grand canonical ensemble}
\label{sec_tl}
In a dilute polymer solution the interaction between $N$
different chains can be neglected so that  
the total free energy of the system is $N$ times the free energy
of a single chain.
We consider the polymer solution within the slit to be in
equilibrium contact with an equivalent polymer solution in a
reservoir outside the slit so that there is exchange of polymer 
coils between the slit and the reservoir. 
The {\it free energy of interaction} between the walls in such a
grand canonical ensemble is given by
\begin{eqnarray}
\delta F
&=& -k_BT\;N\left\{\ln\left(\frac{ {\cal Z}_{||}(D)}{{\cal Z}}\right) -
 \ln\left(\frac{ {\cal Z}_{||}(D=\infty)}{{\cal
Z}}\right)\right\}\;,\nonumber\\
\label{felog}
 & &
\end{eqnarray}
where ${\cal Z}_{||}(D)$ is the partition function of one polymer chain
in a large volume $\cal V$ containing the walls and
${\cal Z}$ is the corresponding partition function of one polymer chain
in the volume $\cal V$ without the walls.
In the thermodynamic limit ${\cal V}\to{\mathbb R}^d$ one has
\cite{ehd,HED,bringer}
\begin{eqnarray}
\label{zb}
{\cal Z} & \to & {\cal V} \hat{{\cal Z}}_b
\end{eqnarray}
with  $\hat{{\cal Z}}_b=\int_{{\mathbb R}^d}d^dr' {\cal Z}_b({\bf r},{\bf r}')$
and where ${\cal Z}_b({\bf r},{\bf r}')$ denotes the partition function of
one polymer chain in the unbounded solution with its ends fixed at
${\bf r}$ and ${\bf r}'$.
Correspondingly ${\cal Z}_{||}({\bf r},{\bf r}')$ denotes the
partition function of one polymer chain within the volume $\cal V$
containing the parallel walls and with its ends fixed at 
${\bf r}$ and ${\bf r}'$.
The volume ${\cal V}={\cal V}_I+{\cal V}_O$ can be divided into the volume
${\cal V}_I$ within the slit and the volume ${\cal V}_O$ outside the
slit. Since the polymer chain cannot penetrate the walls, whose lateral
extensions are large, ${\cal Z}_{||}({\bf r},{\bf r}')$ is nonzero
only if both ${\bf r}$ and ${\bf r}'$ are in ${\cal V}_I$ or in 
${\cal V}_O$ so that
\begin{eqnarray}
{\cal Z}_{||}(D) & = & \int_{\cal V}d^dr \int_{\cal V}d^dr'
{\cal Z}_{||}({\bf r},{\bf r}')
\, =\,  \int_{{\cal V}_O}d^dr \hat{{\cal Z}}_O(z)
+\int_{{\cal V}_I}d^dr \hat{{\cal Z}}_I(z)\quad,
\label{zsep}
\end{eqnarray}
with $\hat{{\cal Z}}_{O,I}(z)=\int_{{\cal V}_{O,I}}d^dr'{\cal
Z}_{||}({\bf r},{\bf r}')$.
In the thermodynamic limit ${\cal V}\to{\mathbb R}^d$ the logarithm
$\ln({\cal Z}_{||}/{\cal Z})$ in Eq.\,(\ref{felog}) 
can be expanded using Eqs.\,(\ref{zb}) and (\ref{zsep}), i.e.,
\begin{eqnarray}
\label{lnzio}
\ln\left(\frac{{\cal Z}_{||}(D)}{{\cal Z}}\right)&=&
\ln\left(1+\frac{{\cal Z}_{||}(D)-{\cal Z}}{{\cal Z}}\right)\\
&\to &\frac{1}{{\cal V}}\frac{{\cal Z}_{||}(D)-{\cal V}\hat{{\cal
Z}}_b}{\hat{{\cal Z}}_b}
\,=\,\frac{1}{{\cal V}}\Bigg[\int_{{\cal V}_O}\!\!\!d^dr
\Bigg(\frac{\hat{{\cal Z}}_O(z)}{\hat{{\cal Z}}_b}-1\Bigg)
+\int_{{\cal V}_I}\!\!\!d^dr
\Bigg(\frac{\hat{{\cal Z}}_I(z)}{\hat{{\cal Z}}_b}-1\Bigg)\Bigg]\;,
\nonumber
\end{eqnarray}
where the ratio $\frac{{\cal Z}_{||}(D)-{\cal Z}}{{\cal Z}}$ is of
the order ${\cal V}_I/{\cal V}$ which tends to zero because the
slit width $D$ is fixed.
Since in the thermodynamic limit the leading contribution to
the first integral of
Eq.\,(\ref{lnzio}) is independent of the slit width $D$ we find for
Eq.\,(\ref{felog})
\begin{eqnarray}
\delta F = -n_{p\,} k_b T & &\left\{\int_{{\cal V}_I}d^dr
\left(\frac{\hat{{\cal Z}}_I(z)}{\hat{{\cal Z}}_b}-1\right)\right.
\label{dftl}
-\left.\left. \int_{{\cal V}_I}d^dr
\left(\frac{\hat{{\cal Z}}_I(z)}{\hat{{\cal Z}}_b}-1\right)
\right|_{D=\infty}\right\}\;,
\end{eqnarray}
with the number density $n_p = N/{\cal V}$ of the polymer chains 
in the bulk solution.
The second integral in Eq.\,(\ref{dftl}) reduces to the sum of
two half-space $(HS)$ integrals which yield contributions
proportional to the area $A$ of the walls\cite{HED}:
\begin{eqnarray}
\left. \int_{{\cal V}_I}\!\!\!d^dr
\left(\frac{\hat{{\cal Z}}_I(z)}{\hat{{\cal Z}}_b}-1\right)
\right|_{D=\infty}
& = & 2 \int_{HS}\!\!\!d^dr\left(\frac{\hat{{\cal Z}}_{HS}(z)}{\hat{{\cal
Z}}_b}-1\right)
\, = \,  - 2 A \frac{\Delta\sigma}{n_p k_B T}\quad,
\label{2fs}
\end{eqnarray}
where we have introduced the surface tension $\Delta\sigma$
between the polymer solution and the confining wall 
[compare Eqs.\,(1.7) and (2.41) in Ref.\,\cite{HED}].
Note that the mean polymer density within 
the slit is determined by the bulk density $n_p$, i.e., in the grand 
canonical ensemble the chemical potential $\mu$ of the polymer coils
is fixed instead of the number $N_I$ of 
polymer coils in the slit (see Subsec.\,\ref{sec_can}).

According to Eqs.\,(\ref{dftl}) and (\ref{2fs}) the total grand
canonical free energy $\Omega$ of the polymer solution
within the slit,
\begin{eqnarray}
\label{fidef}
\Omega &=& - n_{p\,} k_B T A D \omega
\end{eqnarray}
with the dimensionless quantity 
\begin{eqnarray}
\label{doublestar}
\omega &=& \frac{1}{D}\int_0^D dz 
\frac{\hat{{\cal Z}}_I(z)}{\hat{{\cal Z}}_b}\quad,
\end{eqnarray}
can be decomposed as
\begin{mathletters}
\label{pulpe}
\begin{eqnarray}
\label{Gsep}
\frac{\Omega}{n_p k_B T A} &=& D\,f_b+2f_s+\delta f\quad.
\end{eqnarray}
On the rhs of Eq.\,(\ref{Gsep}) appear the reduced bulk free
energy per unit volume
\begin{eqnarray}
f_b &=& -1 \, \, ,
\end{eqnarray}
the reduced surface free energy per unit area 
\begin{eqnarray}
f_s &=& \frac{\Delta\sigma}{n_p k_B T} \, \, ,
\end{eqnarray}
and the reduced free energy of interaction
\begin{eqnarray}
\label{deltaf}
\delta f &=& \frac{\delta F}{n_p k_B T A} \, \, .
\end{eqnarray}
\end{mathletters}
%
\subsection{Canonical ensemble} \label{sec_can}
If instead of the chemical potential $\mu$ the number $N_I$ of polymer 
coils in the slit is used as independent variable, the total free energy $F$ 
of the polymer solution within the slit in the canonical ensemble 
follows from $\Omega$ as the Legendre transform
\begin{eqnarray}
\label{legend}
F(N_I) & = & \Omega(\mu(N_I)) + \mu(N_I)\,N_I\quad,
\end{eqnarray}
where $\Omega$ is given by Eq.\,(\ref{fidef}).
The chemical potential $\mu$ is related to $n_p$ via \cite{plischke}
\begin{eqnarray}
\label{muid}
\mu & = & k_B\,T \ln( n_p \Lambda^d ) \quad,
\end{eqnarray}
where $\Lambda$ is the thermal de Broglie wave length of the
particles, i.e., polymer coils. Equation (\ref{fidef}) implies
for dilute solutions
\begin{eqnarray}
\label{meanN}
N_I & = & -\frac{\partial \Omega(\mu)}{\partial \mu} =
\frac{-n_p}{k_BT}\frac{\partial \Omega}{\partial n_p} =\frac{-\Omega}{k_BT}\,\,.
\end{eqnarray}
Thus $F(N_I)$ is given by
\begin{eqnarray}
\label{fvNi}
F(N_I) & = & -k_BT\,N_I
+k_BT\,N_I\ln\left(\frac{N_I}{A\,D\,\omega} \, \Lambda^d\right)\quad, 
\end{eqnarray}
with $\omega$ from Eq.\,(\ref{doublestar}).

\subsection{Free energy of interaction and mean force}
\label{sec_pma}

We employ the polymer magnet analogy
\cite{gennes2,cloizeaux,schafer,eisen,ehd,HED} in order to calculate
the partition functions $\hat{{\cal Z}}_{HS}(z)$,
$\hat{{\cal Z}}_I(z)$, and $\hat{{\cal Z}}_b$ as needed in
Eqs.\,(\ref{dftl}), (\ref{2fs}), and (\ref{doublestar})
for a single chain with EV interaction.
They are functions of the parameter $u_0$ which characterizes
the strength of the EV interaction and $L_0$ which
determines the number of monomers of the chain 
such that $L_0$ equals ${\cal R}_g^2={\cal R}_x^2\,/\,2$
for an ideal chain in the bulk, i.e., for $u_0 = 0$.
The well-known arguments of the polymer magnet analogy 
\cite{gennes2,cloizeaux,schafer,eisen} imply for the
present case the correspondence
\begin{eqnarray} 
\label{analogy}
{\cal Z}_{||}({\bf r}, {\bf r}' ; L_0, D, u_0) & = &
{\cal L}_{t_0 \to L_0} 
\langle \Phi_1({\bf r}) \Phi_1({\bf r}') \rangle
\Big|_{{\cal N} = 0}
\end{eqnarray}
between ${\cal Z}_{||}({\bf r},{\bf r}')$ and the two-point correlation function 
$\langle \Phi_1({\bf r}) \Phi_1({\bf r}') \rangle$ 
in a ${\cal O}({\cal N})$ symmetric field theory for an 
${\cal N}$-component order parameter field 
${\bf \Phi} = (\Phi_1, \, ... \,, \Phi_{\cal N})$
in the restricted volume ${\cal V}_I$.
In Eq.\,(\ref{analogy}) the operator
\begin{eqnarray} 
\label{laptrafo}
{\cal L}_{t_0 \to L_0} & = &
\frac{1}{2 \pi i} \int_{\cal C} d t_0 \, e^{L_0 t_0}
\end{eqnarray}
acting on the correlation function is an inverse Laplace transform
with ${\cal C}$ a path in the complex $t_{0\,}$-plane to the right
of all singularities of the integrand.
The Laplace-conjugate $t_0$ of $L_0$ and
the excluded volume strength $u_0$ appear, respectively, as the
`temperature' parameter and as the prefactor of the 
$({\bf \Phi}^2)^2$-term in the Ginzburg-Landau Hamiltonian 
\begin{eqnarray}
\label{hamiltonian}
{\cal H} \{ {\bf \Phi} \}
& = & \int_V d^d r  \left\{ \frac{1}{2} (\nabla {\bf \Phi})^2
+ \frac{t_0}{2}  {\bf \Phi}^2 
+ \frac{u_0}{24}  ({\bf \Phi}^2)^2 \right\}
\end{eqnarray}
which provides the statistical weight 
$\exp(- {\cal H} \{ {\bf \Phi} \})$ for the field theory.
The requirement in Eq.\,(\ref{zzero}) describing the repulsive
character of the walls imposes the Dirichlet condition
\begin{equation}
\label{dbc}
{\bf \Phi}({\bf r}) = 0 \, \, , \quad z = 0, D \quad,
\end{equation}
on both walls. This corresponds to the fixed point boundary 
condition of the so-called ordinary transition \cite{binder,diehl}
for the field theory. For the renormalization group improved
perturbative investigations we use a dimensionally regularized 
continuum version of the field theory which we shall 
renormalize by minimal subtraction
of poles in $\varepsilon = 4 - d$ \cite{amit}.
The basic element of the perturbation expansion is the 
Gaussian two-point correlation function (or propagator) 
$\langle \Phi_{i}({\bf r}) \, \Phi_{j}({\bf r}' ) \rangle_{[0]}$
where the subscript $[0]$ denotes $u_0 = 0$ 
[see Eq.\,(\ref{corrfctn}) in Appendix \ref{sec_app}]. 

The loop expansion to first order in $u_0$ and the renormalization
of $\omega$ in Eq.\,(\ref{doublestar}) is completely analogous to that
outlined in Subsec.\,II\,A of Ref.\,\cite{HED}. Some key results
of this procedure relevant for the present case are given in
Appendix \ref{sec_app}.
The final result for $f_s$ and $\delta f$ on the rhs of
Eq.\,(\ref{Gsep}) are given by Eqs.\,(\ref{fseps}) and
(\ref{deltafeps}) in Appendix \ref{sec_app}, where
${\cal N}=0$ for the present polymer
case and $\varepsilon=1$ in $d=3$.

\begin{figure}
\hspace{-0em}
\epsfxsize=13.4cm
\epsfbox{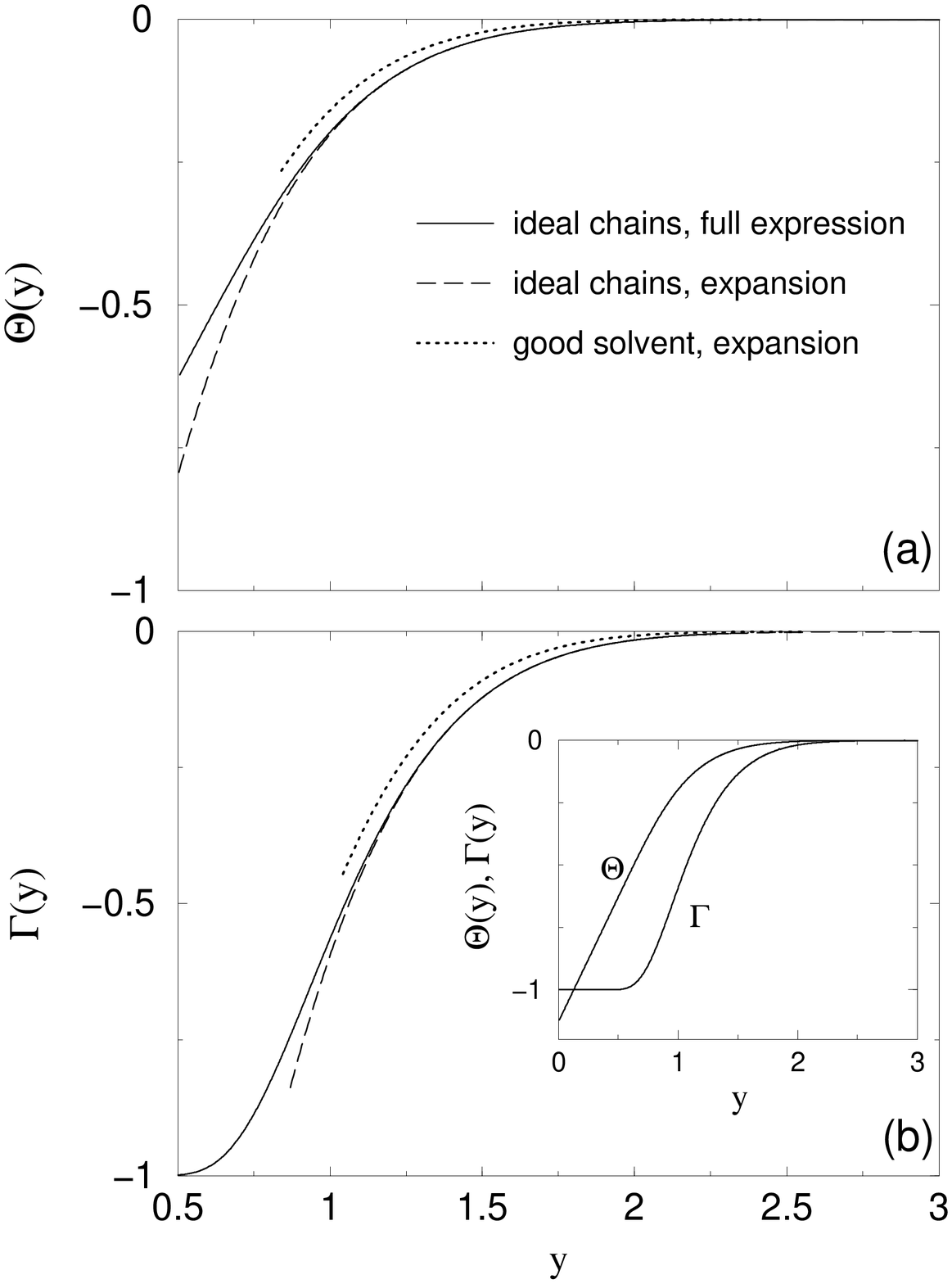}
\caption{
Scaling functions (a) $\Theta(y)$ for the depletion interaction potential
and (b) $\Gamma(y)=-\frac{d\Theta}{d y}$ for the depletion force
between two repulsive, parallel plates 
at a distance $D$ confining a dilute polymer solution in terms of the 
scaling variable $y = D / {\cal R}_x$
[see Eqs.\,(\ref{theta}) and (\ref{Gamma})].
${\cal R}_x=\sqrt{2}\,{\cal R}_g$ for ideal chains and 
${\cal R}_x=1.444\,{\cal R}_g$ in a good solvent for a bulk solution
[see Eqs.\,(\ref{seqD2}) and (\ref{eqD2})] where ${\cal R}_g$ is the
radius of gyration.
For ideal chains the full expression, which is
valid in the whole range of $y$ (solid line; see also the inset), and
an expansion of this
expression, which is valid for $y \gtrsim 1$ (dashed line) are shown.
The same expansion is shown for chains in a good solvent (dotted line).
The dotted line stops where the dashed line starts to deviate
appreciably from the solid line.
}
\label{fig1}
\end{figure}

Figure\,\ref{fig1} shows the universal scaling
function for the free energy of interaction
\begin{eqnarray}
\label{theta}
\Theta(y)=\frac{1}{{\cal R}_x} \, \delta f
\end{eqnarray}
with $\delta f$ from Eq.\,(\ref{deltaf}) and the corresponding
scaling function for the force
\begin{eqnarray}
\label{Gamma}
\Gamma(y)&=&-\frac{\mbox{d}\Theta(y)}{\mbox{d} y}\quad
\end{eqnarray}
in terms of the scaling variable 
\begin{equation} \label{sv}
y = D / {\cal R}_x \, \, .
\end{equation}
Figure\,\ref{fig1} shows both the behavior for ideal chains and
for chains with EV interaction. For chains with
EV interaction the present theoretical approach is only amenable to
capture the behavior for $D/{\cal R}_x$ large, i.e., $y \gg 1$.
As expected the depletion potential and the resulting force is {\em weaker\/} for
chains with EV interaction than for ideal chains, because
 the EV interaction effectively reduces the depletion
effect of the walls.
The inset of Fig.\,\ref{fig1} shows both $\Theta$ and $\Gamma$ 
for ideal chains.
The narrow slit limits read
\begin{eqnarray}
\label{Ty0}
\Theta(y\to 0)&=&-\frac{2}{\sqrt{\pi}}+y
\end{eqnarray}
and
\begin{eqnarray}
\label{Gy0}
\Gamma(y\to 0)&=&-1\quad.
\end{eqnarray}
 Deviations from the linear behavior in Eq.\,(\ref{Ty0}) become
visible only for $y\gtrsim\frac{1}{2}$.
In the opposite limit $y\to\infty$ the leading behavior
for ideal chains is given by
\begin{eqnarray} \label{Tyinf}
\Theta(y\to\infty)&=&-4\sqrt{\frac{2}{\pi}}\frac{1}{y^2}\,e^{-y^2/2}
\end{eqnarray}
and
\begin{eqnarray}\label{Gyinf}
\Gamma(y\to\infty)&=&-4\sqrt{\frac{2}{\pi}}\frac{1}{y}\,e^{-y^2/2}\;.
\end{eqnarray}
The depletion potential in terms of the scaling function
$\Theta(y)$ is attractive. But whether the bulk contribution
$D\,f_b$ has to be taken into account in addition depends on the
'experimental' setup. In the case that the force is measured between
plates immersed in a container filled with the dilute polymer solution
such that solvent and polymer coils
can freely enter the slit from the reservoir, only the scaling
functions $\Theta$ and $\Gamma$ are relevant. This is also
true for the particle wall geometry discussed in Sec.\,\ref{sec_exp}.
But if no exchange is allowed as for the Monte Carlo simulation
discussed in Sec.\,\ref{sec_sim} the force ${\cal K}$ defined
in Eq.\,(\ref{force}) is needed.
%
%
\section{Comparison with Monte Carlo simulations}
\label{sec_sim}
Monte Carlo simulations of polymers are well established both for
ideal chains and for chains with self-avoidance.
In this section we 
compare our results with the simulation of a polymer chain
between two repulsive walls by Milchev and Binder \cite{milchev}
which corresponds to the case studied theoretically here. 
These authors use a bead
spring model for the self-avoiding polymer chain with a
short-ranged repulsive interaction between the beads.
 
In Refs. \cite{unirel} and 
\cite{milchev} (see in particular Fig.$\,4$ in Ref.\,\cite{milchev})
it is stated that
the total force $\cal K$ between the two walls is repulsive and
diverges in the narrow slit limit, i.e.,
\begin{eqnarray}
\label{narrowsl}
\frac{D {\cal K}}{k_B T} & \propto &
\left(\frac{D}{{\cal R}_g}\right)^{-1/\nu}\quad, 
\end{eqnarray}
where $\nu=\frac{1}{2}$ for ideal chains and $\nu=0.588$ for
chains with EV interaction and ${\cal R}_g$ is the radius of 
gyration of the polymer chain in unbounded space \cite{milchev}.
[For the definition of ${\cal R}_g$ see Eqs.\,(\ref{seqD2}) and
(\ref{eqD2}) in Sec.\,\ref{sec_exp}.] 
The qualitative difference to the scaling function $\Gamma(y)$
presented for the force in Subsec.\,\ref{sec_pma} is explained
by the following arguments.
(a) The total force between the walls is repulsive due to the
contribution from the cost in free energy caused by the loss
of the total available space for the polymer chains within
the slit, i.e., the bulk pressure $-f_b$
contributes to the total force.
(b) The total force ${\cal K}$ diverges due to the change from the
grand canonical to the canonical ensemble (see Subsec.\,\ref{sec_can}).
In the simulation in Ref.\,\cite{milchev} the number $N_I$ of polymers 
in the slit is given by one polymer in the slit volume. 
Therefore we obtain the total force ${\cal K}$ 
by differentiating Eq.\,(\ref{fvNi})
with respect to the slit width $D$ and by setting $N_I=1$, i.e., 
\begin{eqnarray}
\label{force}
\frac{D{\cal K}}{k_B T} &=& \frac{1}{\omega}
\frac{\mbox{d}}{\mbox{d} D}\,(D\,\omega)\;,
\end{eqnarray}
where $D\,\omega=-\Omega/(n_{p\,} k_BT\,A)$ is given by the rhs of 
Eq.\,(\ref{Gsep}) in conjunction with Eqs.\,(\ref{fseps}) 
and (\ref{deltafeps}) in Appendix \ref{sec_app}. The lhs of
Eq.\,(\ref{force}) corresponds to the quantity $D f$ in Ref.
\cite{milchev}.

\begin{figure}
\hspace{-1em}
\epsfxsize=15cm
\epsfbox{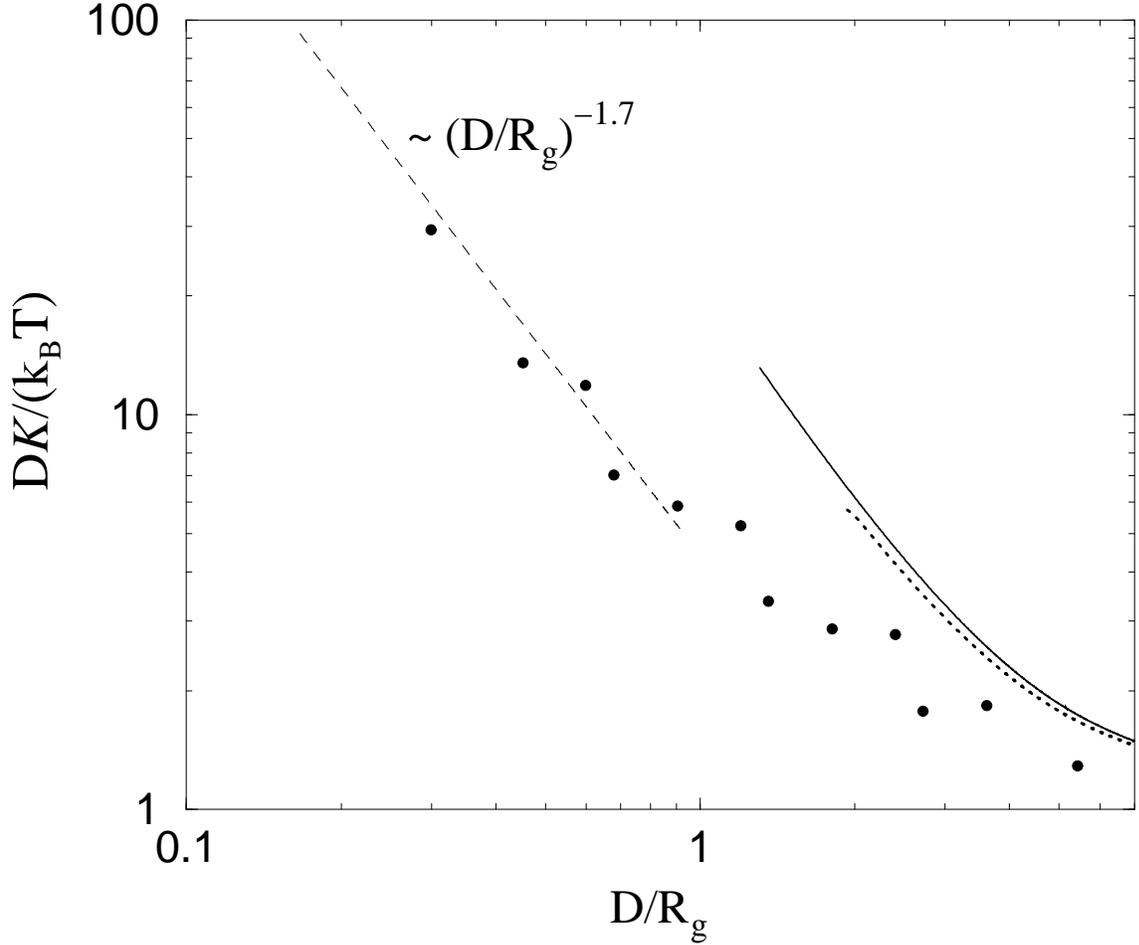}
\caption{
Depletion force ${\cal K}$ [Eq.\,(\ref{force})] between two parallel
walls at distance $D$
confining repulsively a dilute polymer solution. ${\cal R}_g$ is the radius 
of gyration of the chains [see Eqs.\,(\ref{seqD2}) and (\ref{eqD2})]. 
The solid circles correspond to the Monte Carlo simulation data 
in Ref.\,[14]. The force obtained from Eq.\,(\ref{force})
is shown for ideal chains (solid line) and self-avoiding chains
(dotted line). The latter line stops where the expansion for large
$D/{\cal R}_g$ turns out to become unreliable.
The dashed line represents the asymptotic
behavior at small distances for
chains in a good solvent [see Eq.\,(3.1)], with a fit for the
amplitude of the power-law behavior.
For $D/{\cal R}_g \to \infty$ the reduced
force $D{\cal K} / (k_BT)$ tends to $1$.
}
\label{figsim}
\end{figure}

Figure\,\ref{figsim} shows the comparison of
the simulation data of Ref.\,\cite{milchev} and the corresponding 
theoretical result from Eq.\,(\ref{force}), the latter both
for ideal chains and chains with EV interaction. Note that
the theoretical curve for chains with EV interaction is only valid for
large $D/{\cal R}_g$. The curve for chains with EV interaction is
closer to the simulation data than the curve for ideal chains,
in agreement with the fact that the polymer chain in the simulation
is a self-avoiding one. One possible reason for the remaining
deviation might be that the chain in the simulation is too short to
be fully described by the present field-theoretical approach.
The deviation of the Monte Carlo simulation data at large values
of $D/{\cal R}_g$ from the power-law behavior at small values
of $D/{\cal R}_g$ (dashed line) occurs because $\frac{D{\cal K}}{k_B T}$
tends to $1$ for large $D/{\cal R}_g$, which is not captured
by Eq.\,(\ref{narrowsl}).

Figures $5-9$ of Ref.\,\cite{milchev} show density profiles for
the simulated chains with EV interaction, including a comparison
with the analytical result for the profile of ideal chains in the
slit (Fig.\,9 in Ref.\,\cite{milchev}). We want to mention here that
the field-theoretical 
treatment of the monomer density for chains with EV interaction
requires a perturbation expansion involving integrals
over $G*G*G*G$ instead of $G*G*G$ (see Appendix\,\ref{sec_app})
and thus in view of the technical challenges is beyond the scope
of the present study.
Alternative information about the monomer density profiles beyond the
ideal behavior can be found
in Ref.\,\cite{rother} in which a self-consistent mean-field approximation
is used to obtain the monomer density profiles for a single polymer
chain between two repulsive walls. 

\section{Comparison with experiment}
\label{sec_exp}

Rudhardt, Bechinger, and Leiderer \cite{RBL98} have measured
the depletion interaction between a wall and a colloidal 
particle immersed in a dilute solution of nonionic polymer 
chains in a good solvent 
by means of total internal reflection microscopy (TIRM).
They monitored the fluctuations of the relative distance of the
colloid particle from the wall induced by Brownian motion.
From the resulting Boltzmann distribution one can infer
the corresponding effective depletion potential between the
repulsive wall and the particle.

In order to compare these experimental data with our results we
apply the Derjaguin approximation \cite{derjaguin}. 
In the limit that the radius $R$ of the spherical particle
is much larger than both ${\cal R}_x$ and the distance $a$ 
of closest approach surface-to-surface between the particle 
and the wall, the particle can be regarded as composed of
a pile of fringes. Each fringe builds a fringe-like slit with distance
$D = a + r_{||}^2 / (2 R)$, where $r_{||}$ is the radius of the
fringe. Thus the interaction between the particle and the wall
is given by
\begin{eqnarray}
\label{derj}
\frac{\Phi_{\rm depl}(a)}{n_p k_B T} & = & 2 \pi R\, {\cal R}_x^2
\int_0^{\infty} \!\!\! d\upsilon\,\upsilon 
\;\Theta\!\left(\frac{a}{{\cal R}_x}+ \frac{\upsilon^2}{2}\right)\quad, 
\end{eqnarray}
where $\upsilon$ is a dimensionless variable and $\Theta(y)$ is the
scaling function for the free energy of
interaction for the slit geometry [see Eq.\,(\ref{theta})].

\begin{figure}
\hspace{-1em}
\epsfxsize=15cm
\epsfbox{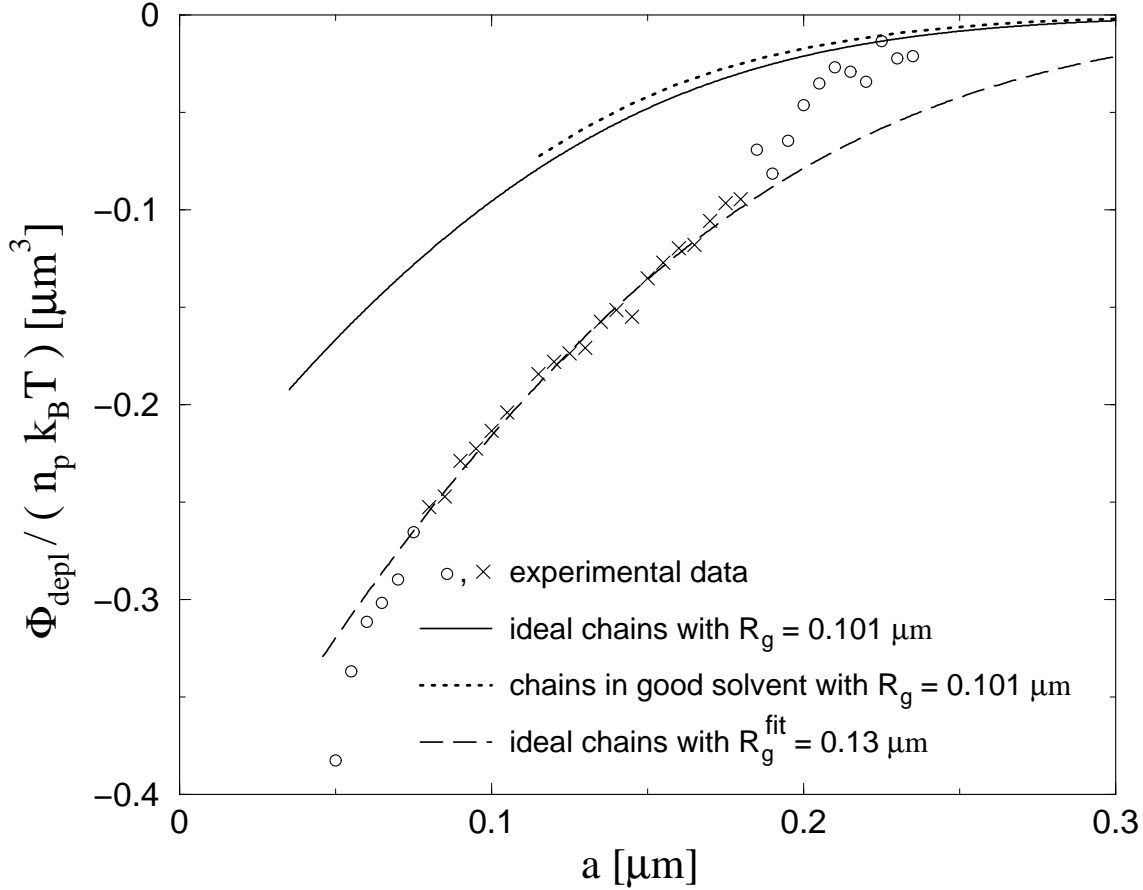}
\caption{
Depletion interaction potential ${\Phi}_{\text{depl}}(a)$ 
between a spherical colloidal particle immersed
in a dilute polymer solution and the container wall as a
function of the distance $a$ of closest approach surface-to-surface
between the sphere and the wall [see Eq.\,(\ref{derj})].
The interaction potential is given in units of $k_BT$ and of the
polymer number density $n_p$.  
The circles and crosses indicate the experimental
data from Ref.\,[6]. Crosses show the range where
the linear relationship $\Phi_{\rm depl}\propto n_p$ allows for a
direct comparison with the theoretical results.
The theoretically calculated depletion interaction
is shown for ideal chains (solid line) and chains in a good
solvent (dotted line) as realized in the experiment.
These curves correspond to the value ${\cal R}_g=0.101\mu{\rm m}$,
which has been determined by independent experiments
[see Eqs.\,(\ref{seqD2}) and (\ref{eqD2})].
Using the radius of gyration as a fit parameter yields the dashed line
corresponding to ${\cal R}_g=0.13\mu{\rm m}$ and ideal chains.
} 
\label{figexp}
\end{figure}

In Ref.\,\cite{RBL98} the interaction potential $\Phi_{\rm depl}(a)$
is measured for nonionic polymer chains in a good solvent for
polymer number densities 
$n_p=0$, $7.6$, $10.2$, $12.7$, and $25.5 \, \mu \mbox{\rm m}^{-3}$. 
All these polymer number densities
represent a {\em dilute\/} polymer solution so that 
$\Phi_{\rm depl}(a)$ is a linear function of $n_p$ [see Eq.\,(\ref{felog})].
Therefore our results are applicable and Fig.\,\ref{figexp} shows
$\Phi_{\rm depl}/n_p$ as a function of $a$.
The crosses in Fig.\,\ref{figexp} correspond to those values of $a$ for
which the above mentioned linear behavior $\Phi_{\rm depl}\propto n_p$
(not shown) is in good agreement with the experimental data.
Deviations from the linear behavior $\Phi_{\rm depl}\propto n_p$
for small and large particle-wall distances $a$ (not shown) can be explained
by the experimental method TIRM used in Ref.\,\cite{RBL98}:
a higher interaction potential implies a lower probability
to find the particle at the corresponding distance causing
lower accuracy. It turns out that the total interaction potential as
the sum of depletion potential, electrostatic repulsion, and gravity
is highest for short and large distances, which are those where the
linear relationship $\Phi_{\rm depl}\propto n_p$ is not confirmed
experimentally. Figure\,\ref{figexp} also shows the corresponding theoretical
predictions for the experimental data, both for ideal polymer chains
and chains with EV interaction, i.e., chains in a good solvent
as realized in the experiment. Note that all parameters entering
these theoretical predictions are {\em fixed} by available
experimental data for $n_p$, $a$, and ${\cal R}_g$, i.e., there are no
freely adjustable parameters. In particular, the radius of gyration
${\cal R}_g$ of the polymer chains used in Ref.\,\cite{RBL98} has been
measured fairly accurately by means of small angle scattering of
x-rays \cite{selser}, resulting in ${\cal R}_g=0.101 \mu\mbox{\rm m}$.
According to the definition of ${\cal R}_g$ as measured in
small angle scattering experiments \cite{scatter}, i.e.,
\begin{equation}
\label{seqD2}
{\cal R}_g^2= 
\frac{\int d^3r\int d^3r'\rho({\bf r})\rho({\bf r}')|{\bf r}-{\bf r}'|^2}
{2\int d^3r\int d^3r'\rho({\bf r})\rho({\bf r}')} \, \, , 
\end{equation}
where $\rho({\bf r})$ is the monomer density,
in $d=3$ one has \cite{grass}
\begin{equation}
\label{eqD2}
{\cal R}_g= 0.6927 \, {\cal R}_x \, \, , \quad d = 3 \, \, .
\end{equation}

Figure\,\ref{figexp} shows that the experimental data of 
Ref.\,\cite{RBL98} deviate from the theoretical result 
derived here.
Note that this deviation is not visible if the polymer size is used
as a freely adjustable parameter in order to gain agreement between
the experimental and theoretical data. The theoretical curve for
chains with EV interaction, as realized in the experiment, is even
{\em further away} from the experimental data than the theoretical
curve for ideal chains.
The latter observation 
confirms the necessity of reinterpreting of the experimental
data, e.g., by adjusting the radius of gyration. This can be done
such that agreement with the theoretical data for chains with
EV interaction is gained for the intermediate values of the distance
where the linear relationship $\Phi_{\rm depl}\propto n_p$ allows for a
direct comparison with the theory.
[The remaining differences outside this intermediate regime ($\circ$)
pose a separate issue as discussed in the paragraph following
Eq.\,(\ref{derj}).]
However the difference between the two theoretical
curves for ideal chains and chains with EV interaction is small
compared to the deviation from the experimental data.

Any attempt to use the Monte Carlo data
obtained in Ref.\,\cite{milchev} for predicting the depletion interaction
between a colloidal particle and a wall would require
two integrations of ${\cal K}$ and a change to the grand canonical
ensemble, which poses prohibitive accuracy problems. Nonetheless a
qualitative statement can be given easily.  
Figure\,\ref{figsim} shows that the force obtained in the Monte
Carlo simulation is weaker than the force calculated
for chains with EV interaction. This is also expected to hold for the
depletion interaction between a particle and the wall.

\section{Concluding remarks and summary}
\label{sec_con}
Based on field-theoretical techniques we have determined the effective
depletion interaction between two non-adsorbing walls confining a
dilute solution of long flexible polymer chains.
Our main results are:
\begin{enumerate}
\item{The field-theoretical calculation yields the universal scaling
functions of the depletion interaction potential and the 
corresponding force for ideal chains and 
for chains with excluded volume interaction in the limit
$y = D / {\cal R}_x \gg 1$ , where $D$ is the separation between
the walls and ${\cal R}_x$ is the projected end-to-end distance 
of the chains [see Eqs.\,(\ref{theta}), (\ref{Gamma}), and
(\ref{eqD2}) and Fig.\,\ref{fig1}]. 
The depletion potential is weaker for
chains in a good solvent than for ideal chains.}
\item{For $y \gtrsim 1$ we find fair agreement with corresponding
Monte Carlo simulation data \cite{milchev} if the excluded volume
interaction is taken into account (see Fig.\,\ref{figsim}).
We surmise that remaining
discrepancies are due to higher order terms in the field-theoretical
calculation which are not yet taken into account and due to the
possibility that the length of the simulated polymer chain has not yet
reached the scaling limit for which the field theory is valid.}
\item{Using the Derjaguin approximation we have compared our
theoretical results with the experimental data \cite{RBL98} for the
depletion potential between a spherical colloidal particle and a wall
(see Fig.\,\ref{figexp}). 
We obtain a fair agreement only if the radius of gyration ${\cal R}_g$
of the polymers is used as a fit parameter. This value, however,
differs from independently determined values for ${\cal R}_g$. The
reasons for these differences are not yet understood. The excluded
volume interaction between the monomers of the polymer chain plays
only a minor role for reaching agreement between theory and experiment.}
\end{enumerate}

\section*{Acknowledgments}

We thank A. Milchev and K. Binder for a helpful
discussion of their simulation data and
C. Bechinger for providing the experimental data.
This work has been supported by the 
German Science Foundation both through
Sonderforschungsbereich 237 
{\em Unordnung und gro{\ss}e Fluktuationen\/}
and the graduate college {\em Feldtheoretische und numerische Methoden
der Elementarteilchen- und Statistischen Physik\/}.

\appendix
\section{}
\label{sec_app}
The Gaussian two-point correlation function in the slit of width $D$ reads
\begin{eqnarray}
\label{corrfctn}
\langle \Phi_{i}({\bf r}) \, \Phi_{j}({\bf r}' ) \rangle_{[0]} 
\, & = & \, \delta_{ij} \, G({\bf r}, {\bf r}'; t_{0\,}, D)
 = \delta_{ij} \, \widehat{G} 
(| {\bf r}_{\parallel} - {\bf r}_{\parallel}'
|, z , z\,' \, ; t_{0\,}, D) \\
& = & \delta_{ij}\int\frac{d^{d-1}p}{(2\pi)^{d-1}} \exp[i{\bf
p}\cdot({\bf r}_{\parallel} - {\bf r}_{\parallel}')]
\, \widetilde{G} ({\bf p}, z, z'; t_0, D)\nonumber\quad,
\end{eqnarray}
with the Gaussian propagator $\widetilde{G} ({\bf p}, z, z'; t_0,
D)$ in $p$-$z$ representation given by \cite{diehl,KED95} 
\begin{eqnarray}
\label{filmprop}
\widetilde G({\bf p}, z, z'; t_0, D) & = & \frac{1}{2 b}
\bigg[ \, e^{-b|z-z'|} - e^{-b(z+z')} \\
&&\quad + \, \frac{e^{-b(z-z')} +e^{-b(z'-z)}-e^{-b(z+z')} - e^{b(z+z')}}
{e^{2bD}-1}\bigg] \;,\nonumber\\
\end{eqnarray}
where $b=\sqrt{p^2+t_0}$.
The loop expansion of
the total susceptibility reads \cite{klimpel}
\begin{eqnarray}
\label{chiloop}
\chi(t_0, D, u_0) & = & \int_0^D \!\!\! dz  dz' \widetilde{G}({\bf 0},z,z')
 \\
&& -  \frac{u_0}{2}\frac{{\cal N}+2}{3} \int_0^D \!\!\! dz
 dz'  dz'' \int\frac{d^{d-1}p}{(2\pi)^{d-1}}\,
\widetilde{G}({\bf 0},z,z'')
\,\widetilde{G}({\bf p},z'',z'')
\,\widetilde{G}({\bf 0},z'',z')
\nonumber\\[2mm]
&& + \, O(u_0^2)\nonumber\quad.
\end{eqnarray}
The procedure outlined in Subsec.\,II\,A of Ref.\,\cite{HED}
yields the renormalized total susceptibility in one loop order as
(see also Appendix B of Ref.\,\cite{klimpel})
\begin{eqnarray}
\label{loopchiren}
\frac{\chi_{\rm ren}(\tau=(D\mu)^2 t, D, u)}{ D^3}
& & = \, \frac{1}{\tau} - \frac{2}{\tau^{3/2}}
  \frac{1-e^{-\sqrt{\tau}}}{1+e^{-\sqrt{\tau}}}\\
& & +\,u\,\frac{{\cal N}+2}{3} \frac{2}{\tau^{3/2}}
\Bigg\{ \bigg[ \sqrt{\tau} + 2 \sqrt{\tau}
  \frac{e^{-\sqrt{\tau}}}{(1+e^{-\sqrt{\tau}})^2} - 3
  \frac{1-e^{\sqrt{\tau}}}{1+e^{\sqrt{\tau}}} \bigg] \nonumber \\
& &\;\times \, \bigg[ 2 f_1 + 2 \ln(D\mu) - \ln \tau + \ln(4\pi)  + 1 -
  C_E 
 - \, 8 \int_1^\infty\!\!\! d s \frac{\sqrt{s^2-1}}{e^{2 \sqrt{\tau}
  s}-1} \bigg] \nonumber \\[1mm]
& &\; - \,4 \pi \, \frac{\frac{1}{2}-\frac{1}{\sqrt{3}} +
  e^{-\sqrt{\tau}}(2-\sqrt{3}) + e^{-2\sqrt{\tau}}(\frac{1}{2}-\frac{1}{\sqrt{3}})
  }{(1+e^{-\sqrt{\tau}})^2} \nonumber \\[1mm]
& &\;- \, 4 \, \frac{1-e^{-\sqrt{\tau}}}{1+e^{-\sqrt{\tau}}}\int_0^\infty\!\!\! d s
  \frac{\sqrt{s^2-1}}{e^{2\sqrt{\tau}s}-1}
\, \Bigg( \frac{2}{s+\frac{1}{2}} +
  \frac{2}{s-\frac{1}{2}} + \frac{1}{s-1} - \frac{1}{s+1} \, \Bigg)
\;\Bigg\}\nonumber\\
&& +\,O(u^2)  \nonumber\;. 
\end{eqnarray}
Here $\mu$ is the inverse length scale which determines the
renormalization group flow; $t$ and $u$ are the renormalized and
dimensionless counterparts of $t_0$ and $u_0$, respectively;
$C_E$ is Euler's constant and for the definition of the constant $f_1$
we refer to Ref.\,\cite{HED}. 

The free energy is obtained via the inverse Laplace transform
of $\chi_{\rm ren}(D)$ and normalization by the transformed
renormalized
total susceptibility for the unbounded space $\chi_{{\rm ren},b}$.
The decomposition into bulk, surface, and finite-size contributions
is carried out by the analysis of the scaling behavior of these
parts of the free energy.
The surface and finite-size parts of the free energy
[see Eq.\,(\ref{pulpe})] at the fixed point 
of the renormalization group
and for ${\cal N}=0$ with the scaling variable
$y=D/{\cal R}_x$ are given by
\begin{eqnarray}
\label{fseps}
f_s & = & {\cal
R}_x\,\sqrt{\frac{2}{\pi}}\,\left\{1-\frac{\varepsilon}{4}\left[1-\frac{3\ln
2}{2}-\frac{\pi}{2}+\frac{\pi}{\sqrt{3}}\right]\right\}\quad,
\end{eqnarray}
and
\begin{eqnarray}
\label{deltafeps}
&&\frac{\delta f}{D}  =  4 \,\mbox{Erfc}\Big( \frac{y}{\sqrt{2}} \Big) -
4\,\sqrt{\frac{2}{\pi}}\frac{1}{y}\, e^{-y^2/2} \\[1mm]
&&-\frac{\varepsilon}{4}\Bigg[\frac{e^{-y^2/2}}{y\sqrt{2\pi}}
 \Big( 4 + \frac{4 \pi}{\sqrt{3}} -4 \pi + 6 \ln (2y^2) 
- 6 C_E \Big) 
+\mbox{Erfc}\Big( \frac{y}{\sqrt{2}} \Big)
 \Big( 2 \pi - \frac{2 \pi}{\sqrt{3}} - 2 \ln (2y^2)
+ 2 C_E \Big) \nonumber \\[1mm]
&&\qquad- {\cal L}_{\tau \to \frac{1}{2y^2}} \Big\{
\frac{e^{-\sqrt{\tau}} \ln \tau}{\tau} \Big\}
-3\,{\cal L}_{\tau \to \frac{1}{2y^2}} \Big\{ \frac{e^{-\sqrt{\tau}} \ln
  \tau}{\tau^{3/2}} \Big\} \Bigg]
\;+\; O (\varepsilon^2) \nonumber \quad,
\end{eqnarray}
where $\chi_{\rm ren}$ in Eq.\,(\ref{loopchiren}) is expanded for large
plate separations $D$ up to order $O(e^{-\sqrt{D^2t}})$, because for
small distances the correct behavior including the dimensional crossover cannot be
obtained even for the full expression. But using the expansion
has the additional benefit to yield partly analytical results
for the separations $D$ of interest here. 
For $d = 3$,
$R_x$ is related to the radius of gyration ${\cal R}_g$ by Eq.\,(\ref{eqD2}). 
The full result for ideal chains for $\delta f$ is given by
Eq.\,(\ref{loopchiren}) for $u=0$, i.e.,
\begin{eqnarray}
\label{fullmf}
\delta f & = & -D {\cal L}_{\tau\to\frac{1}{2y^2}} \left\{ \frac{4}{\tau^{3/2}}
\frac{1}{1+e^{\sqrt{\tau}}} \right\} \quad, 
\end{eqnarray}
which is valid for all $y$.


\end{document}